\documentclass[superscriptaddress, nofootinbib, twocolumn, amsmath,amssymb, aps, pra, notitlepage, longbibliography]{revtex4-1}

\usepackage{dcolumn}
\usepackage{bm}
\usepackage{epsfig}
\usepackage{graphicx}
\usepackage{latexsym}
\usepackage{amsfonts}
\usepackage{setspace}
\usepackage{graphicx}
\usepackage{amsmath}
\usepackage{verbatim}
\usepackage{color}
\usepackage{SIunits}
\usepackage{hyperref}
\usepackage{nccmath}

\begin{document}

\title{A chip-integrated coherent photonic-phononic memory}

\author{Moritz Merklein$^{1,2,\ast}$, Birgit Stiller$^{1,2,\ast,\dagger}$, Khu Vu$^{3}$,  Stephen J. Madden$^{3}$ and Benjamin J. Eggleton$^{1,2}$\\
\small{ \textcolor{white}{blanc\\}
$^{1}$Centre for Ultrahigh bandwidth Devices for Optical Systems (CUDOS), Institute of Photonics and Optical Science (IPOS), School of Physics, University of Sydney, Sydney, New South Wales 2006, Australia.\\
$^{2}$Australian Institute for Nanoscale Science and Technology (AINST), University of Sydney, Sydney NSW 2006, Australia.\\
$^{3}$Centre for Ultrahigh bandwidth Devices for Optical Systems (CUDOS), Laser Physics Centre, Research School of Physics and Engineering, Australian National University, Canberra, Australian Capital Terrritory 0200, Australia.\\
$^{\ast}$These authors contributed equally to this work.\\
$^{\dagger}$Corresponding author: birgit.stiller@sydney.edu.au}}

\begin{abstract} 

  Controlling and manipulating quanta of coherent acoustic vibrations - phonons - in integrated circuits has recently drawn a lot of attention, since phonons can function as unique links between radiofrequency and optical signals, allow access to quantum regimes and offer advanced signal processing capabilities. Recent approaches based on optomechanical resonators have achieved impressive quality factors allowing for storage of optical signals. However, so far these techniques have been limited in bandwidth and are incompatible with multi-wavelength operation. In this work, we experimentally demonstrate a coherent buffer in an integrated planar optical waveguide by transferring the optical information coherently to an acoustic hypersound wave. Optical information is extracted using the reverse process. These hypersound phonons have similar wavelengths as the optical photons but travel at 5-orders of magnitude lower velocity. We demonstrate the storage of phase and amplitude of optical information with GHz-bandwidth and show operation at separate wavelengths with negligible cross-talk.

\end{abstract}

\maketitle

Storing or delaying optical signals has been a major driving force for a wide variety of research efforts as it offers new possibilities in all-optical processing and enhanced light matter interactions. An optical buffer that is able to maintain the coherence of the optial signal, i.e. storing amplitude and phase information, and is able to operate at multiple wavelengths would greatly enhance the capacity of photonic integrated circuits and optical interconnects. Coupling light to coherent acoustic phonons in optomechanical systems offers not only the opportunity to slow down the velocity of an optical pulse \cite{Weis2010,Safavi-Naeini2011}, but also enables a full transfer of an optical wave to an acoustic wave\cite{Verhagen2012,Fiore2011a,Galland2014,Wang2013b}, which subsequently can be transferred back to the optical domain after a certain storage time. Recent years have seen great progress in increasing the storage time in photonic-phononic whispering gallery mode resonators\cite{Dong2015,Kim2015,Shen2016,Fiore2011a} and optomechanical cavities\cite{Fang2016a,Balram2015,Li2015a}, with reported storage times in the order of microseconds. Furthermore, the photon-phonon-photon transfer can be fully coherent\cite{Verhagen2012,Balram2015,Fiore2013}. However, there are several major challenges which need to be addressed before an optical memory based on this approach is compatible with all-optical information processing and transmission techniques.

\begin{figure*}[!t]
\begin{center}
  \includegraphics[width=1.5\columnwidth]{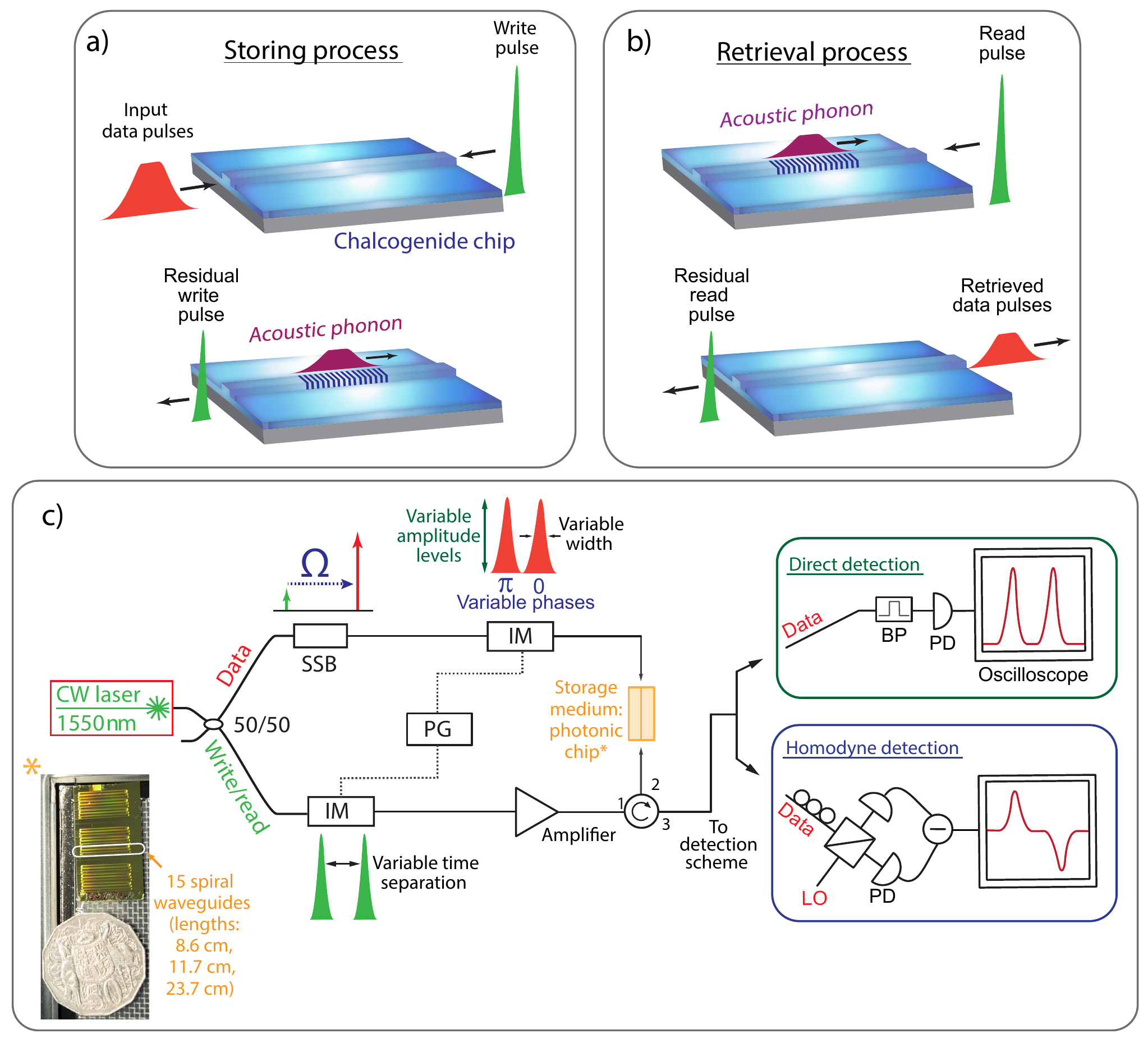}
\caption{\textbf{Basic principle and set-up of the photonic-phononic memory} \small{a) An optical data pulse is depleted by a strong counter-propagating write pulse, storing the data pulse as an acoustic phonon. b) In the retrieval process a read pulse depletes the acoustic wave, converting the data pulse back into the optical domain. c) A basic scheme of the experimental set-up. The inset shows a chalcogenide chip next to a 50-cent coin. The chip contains more than 100 spiral waveguides with different lengths. Note: This is only a scheme and the actual set-up is more advanced and can be found in the supplementary material (CW: continuous wave; SSB: single-sideband modulator; IM: intensity modulator; PG: pulse generator; BP: bandpass filter; PD: photo-detector; LO: local oscillator.)} 
}
\label{princ}
\end{center}
\end{figure*}

First, any practical optical buffer needs, amongst other requirements, at least a GHz bandwidth. Previous demonstrations relied on structural resonances – either in the form of high-Q resonators or suspended optomechanical cavities, in which the bandwidth is limited to sub-MHz. There are several theoretical proposals to increase the bandwidth in optomechanical systems\cite{Safavi-Naeini2011b,Chang2010a} but there has been no experimental demonstration to date. \\Second, optical data transmission schemes usually harness multiple wavelength channels to increase the overall capacity. This means the storage process needs to work over a wide frequency range (a large number of channels); and it needs to preserve the frequency of the optical signal (no cross-talk between the channels). These requirements are challenging to fulfill in photonic-phononic systems relying on structural resonances, e.g. silica fiber-tip whispering gallery mode resonators or photonic-phononic crystal defect modes, since the optical wavelength is strictly bound to the resonance frequency. In the case of whispering gallery mode resonators, an optical pulse transferred to a phonon can be retrieved by a read pulse at a different wavelength \cite{Fiore2011a}, so several wavelength channels cannot be stored and retrieved unambiguously, since the storage/retrieval process is not frequency preserving\cite{Hill2012,Fan2016}.\\
Finally, a practical optical buffer must be chip-integrable and able to be interfaced with other on-chip components, criteria not easily satisfied with other optomechanical platforms investigated to date. Fused silica fiber-tip resonators are micrometre size \cite{Dong2015,Kim2015,Shen2016,Fiore2011a,Fiore2013} but cannot be easily implemented onto a planar chip platform. Lithographically-produced photonic-phononic crystals, which form resonant cavities for the acoustic and optical modes, possess the requisite small footprint\cite{Fang2016a,Balram2015}. Despite this, they either rely on fiber taper coupling\cite{MingCaiOskarPainter2000,Vahala2003} that can be challenging to operate outside a laboratory environment or require complicated under-etching processes to maximize the optical and acoustic Q-factor\cite{Safavi-Naeini2010}. The underetching step is required to confine the optical and acoustic modes in the out-of-plane direction and avoid leakage to the substrate, but limits compatibility with planar integrated photonic circuits.\\
Here we demonstrate a different approach for coherent optical storage, harnessing traveling acoustic phonons in a planar integrated waveguide. We transfer the information carried by the optical signal to these acoustic phonons using stimulated Brillouin scattering (SBS)\cite{Boyd2003,Zhu2007}. We demonstrate that this transfer is fully coherent by storing and retrieving different phases. Our buffer does not rely on a structural resonance, so is not limited to a narrow bandwidth or single wavelength operation. We show that the unique phase matching condition between traveling acoustic and optical waves allows the unambiguous storage and retrieval at several different wavelengths without cross-talk. \\
This result was enabled by a recent paradigm shift in SBS research from long lengths of optical fiber to chip-scale devices, allowing the excitation of coherent acoustic phonons on a chip using optical forces\cite{Pant2011,VanLaer2015,Kittlaus2015,Merklein2016a,Eggleton2013}. The opto-acoustic interaction strength is increased by several orders of magnitude by using carefully designing waveguides that guide optical as well as acoustic waves, allowing us to store broadband optical signals in a planar waveguide without relying on a resonator geometry. The acoustic phonons travel in the waveguide at a velocity that is 5-orders of magnitude slower than in the optical domain and do not suffer from effects of optical dispersion and other detrimental optical non-linearities during the delay process. Transferring the signal back to the optical domain leads to a delay of the optical signal by approximately the time the signal was encoded as acoustic wave.\\
In this article we exploit this ultra-strong local opto-acoustic interaction in a highly nonlinear chalcogenide spiral waveguide to demonstrate for the first time storage of several optical bits with sub-ns pulse-width corresponding to a broad GHz bandwidth. We show the retrieval of the phase and amplitude information, multi-wavelength operation and continuously adjustable storage time over 21 pulse widths. We confirm our measurements using simulations based on coupled-mode equations showing excellent agreement with the experimental results. Our photonic-phononic memory operates at room temperature and only relies on a planar waveguide that can be interfaced with other on-chip components in a straight forward manner. \newline

\subsection{Results} 

\begin{figure*}[!htb]
  \centering
  \includegraphics[width=2\columnwidth]{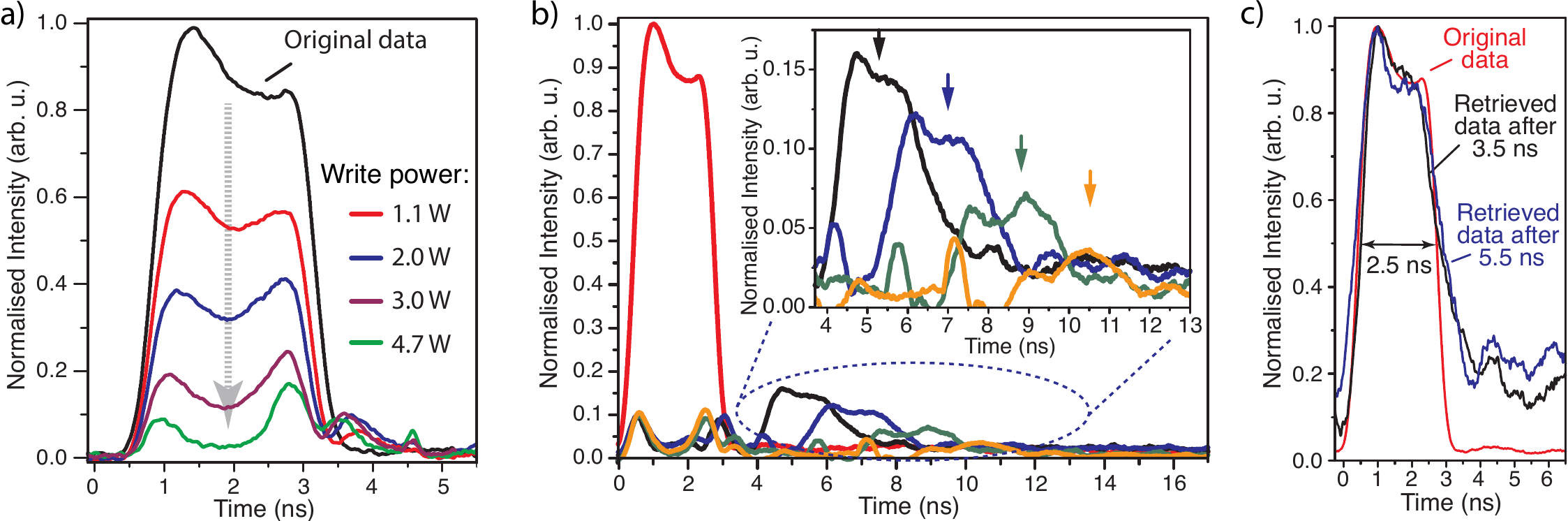}
\caption{\textbf{Store and retrieval process with tunable storage time} \small{a) Storing process: The optical data pulse is depleted by the counter-propagating write pulse, transferring the carried information to the acoustic phonon. Depletion of more than 90\% can be achieved. b) Retrieved data pulses after different storage times. The inset shows a zoomed-in version of the retrieved data pulses. c) Original data pulse super-imposed with the retrieved data pulses after 3.5\,ns and 5.5\,ns respectively; the details of the original shape can be distinguished in the retrieved data pulses.}}
\label{result}
\end{figure*}

We use SBS - one of the strongest nonlinear effects - to coherently couple two optical waves and an acoustic wave. The optical data signal is transferred to the acoustic wave by a strong counter-propagating optical write signal. Using this nonlinear effect as a memory was first proposed in highly nonlinear fiber\cite{Zhu2007,Santagiustina2013}. The storage of only one amplitude level of several nano-second long pulses has been shown to date\cite{Zhu2007}. This is only a fraction of the capability this memory concept can offer.\newline
Our approach to write and retrieve the optical data pulses as acoustic phonons is schematically shown in figure \ref{princ}a) and \ref{princ}b). A strong optical write pulse $\omega_{write}$, offset by the acoustic resonance frequency of the optical waveguide material, propagates counter to the optical data pulse $\omega_{data}$. When the two pulses encounter each other, the beat pattern between the two compresses the material periodically through a process known as electrostriction, exciting resonantly and locally a coherent acoustic phonon $\Omega = \omega_{data} - \omega_{write}$. The required power depends on the local Brillouin gain, which is orders of magnitude higher in chalcogenide $As_{2}S_{3}$ rib waveguides than, for example, in standard silica fiber. The optical and the acoustic modes are guided in the rib waveguide structure by the refractive index contrast and the acoustic impedance between the chalcogenide glass and the silica surrounding, respectively\cite{Poulton2013a}. Once transferred to the acoustic wave the information on the optical data pulses can be retrieved after a storage time of several nanoseconds, corresponding to several tens of data pulse widths. The process is shown in figure \ref{princ}b). A strong read pulse is coupled into the waveguide and retrieves the optical information by depleting the acoustic wave, the inverse process of the writing step. The set-up for the photonic buffer is schematically shown in figure \ref{princ}c) (a detailed description of the setup can be found in the methods section and the supplementary material).

As a storage medium we use a small footprint spiral waveguide made out of the chalcogenide glass $As_{2}S_{3}$ comprising a rib waveguide structure with a cross-section of 2.2 $\mathrm{\mu}$m by 800\,nm. A photo of the chip is depicted in the inset of figure \ref{princ} c) next to an Australian 50-cent coin. Every chip consists of spirals with several lengths ranging from around 9\,cm to 24\,cm. The spiral waveguides are grouped in quintets with a footprint per group of 20\,$\times$\,0.7\,mm. Longer waveguides are available by repeating the same spirals on one chip during the fabrication, leading to waveguides with up to 46\, cm length. For details on the fabrication methods of the chip we refer to reference\cite{Madden2007}. Lensed fiber-tips are used to couple light in and out of the waveguides. The chalcogenide glass is sandwiched between the silica substrate and the silica over-cladding. This not only provides guidance of the optical mode due to a contrast in the refractive index but provides also an acoustic impedance mismatch between the soft chalcogenide glass ($v_\mathrm{sound}$=2500\,m/s) and the stiff silica ($v_\mathrm{sound}$=5996\,m/s). Both the optical and the acoustic waves are guided in the chalcogenide glass, which provides a large opto-acoustic overlap. Ultra-high Brillouin gain of up to 50\,dB amplification of a small CW seed for a moderate CW pump power of 300\,mW was achieved.\\
The experimental realizations of an all-integrated multi-wavelength coherent photonic-phononic buffer is shown in figures \ref{result}, \ref{result2} and \ref{multi}. Figure \ref{result}a) shows the depletion of the optical data pulse with increasing counter-propagating write pulse power (storing process). For this experiment the storage medium is a 46\,cm long spiral waveguide. Due to the ultra strong Brillouin gain in chalcogenide waveguides, the depletion reaches over 90\% with 20-fold lower write pulse peak power than in highly nonlinear fiber approaches\cite{Zhu2007} for similar pulse conditions. The peak power levels of the interacting optical pulses presented in this article vary from 10\,mW to 50\,mW for the data pulses and 3\,W to 10\,W for the write and read pulses depending on the overall gain of the individual waveguides.

The storage and subsequent retrieval of the optical data pulses are demonstrated in figure \ref{result}b). The storage time can be continuously adjusted by simply controlling the time difference between the read and the write pulses. A readout efficiency of 15\% to 32\% after 3.5\,ns was achieved (see also supplementary). The inset of figure \ref{result}b) shows a zoomed-in version of four examples of retrieved data pulses after different storage times. From the exponential decrease of the retrieval efficiency an acoustic decay time of 10.5\,ns is measured using an exponential fit and is confirmed by a pump-probe measurement of the Brillouin gain linewidth (see supplementary material). In order to study the retrieval of the pulse shape, we superimpose the normalized original data pulse with two normalized retrieved data pulses, displayed in figure \ref{result}c). The shape of the optical data pulse is maintained during the storage process indicating that the bandwidth of the photonic-phononic memory is large enough to resolve even small features, such as the peak at the beginning of the optical data pulse. The intrinsic Brillouin linewidth is only in the range of tens of MHz, however due to the strong opto-acoustic coupling in the photonic-phononic waveguides the Brillouin response can be broadened to several GHz\cite{Zhu2011}. \newline \indent

\subsection{Phase coherence and multiple amplitude storage}

\begin{figure*}[!htb]
  \centering
  \includegraphics[width=1.5\columnwidth]{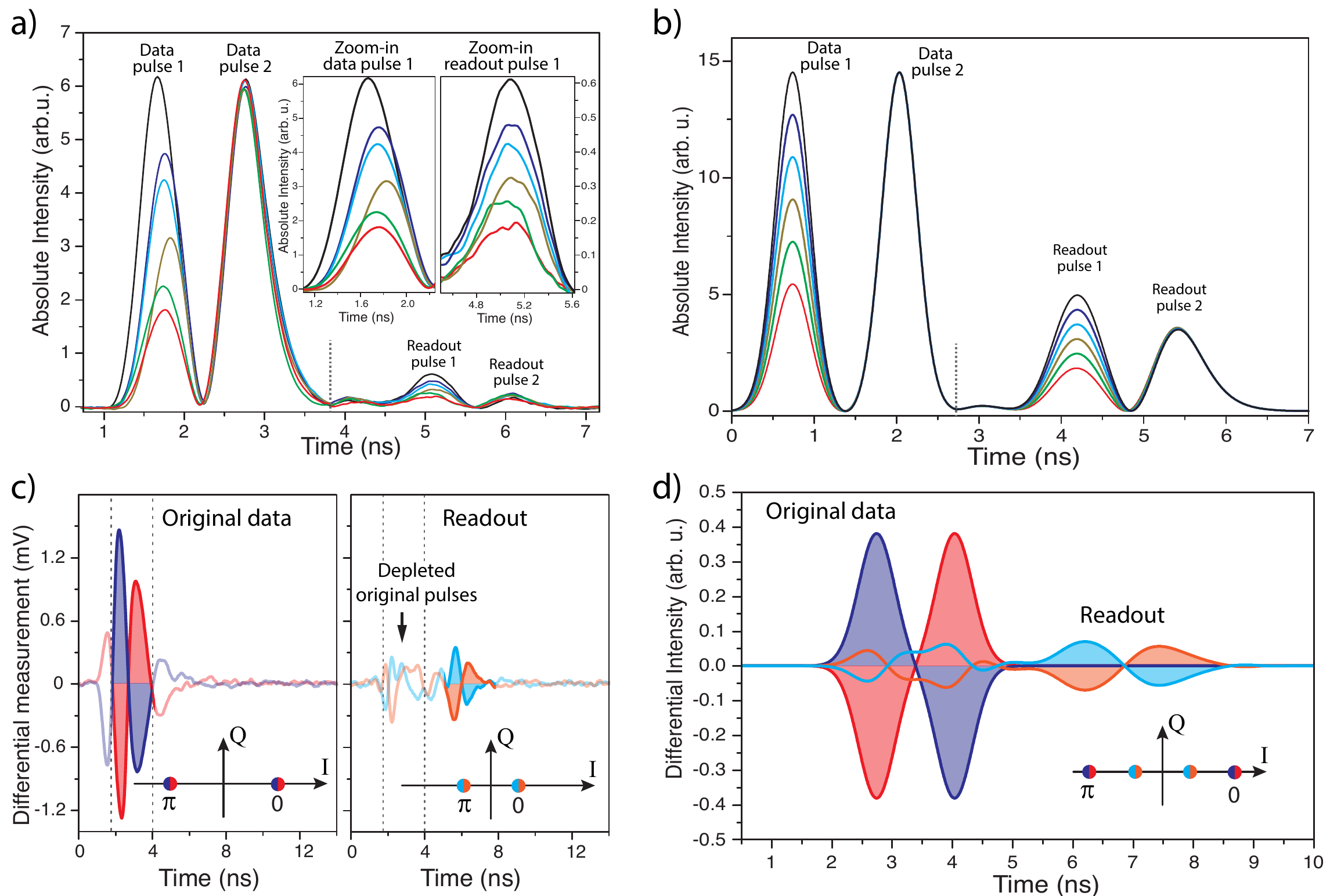}
\caption{\textbf{Amplitude and phase encoded signals} \small{a) Six different amplitude levels of a 500\,ps optical pulse can be stored and retrieved after a storage time of 3.5\,ns via direct detection. The amplitude of pulse 2 remains constant for the original and retrieved data pulse. The read-out efficiency of pulse 2 is lower due to practical limitations in the experiment (power limitation). b) Simulation data of the amplitude response of the system. c) Two phase levels of two 500\,ps optical pulses, either 0 and $\pi$ or $\pi$ and 0 are retrieved via homodyne detection after a storage time of 3.5\,ns. The lower inset shows the two phases in phase space. d) Simulated phase response of the system.}}
\label{result2}
\end{figure*}

We showed in the previous section that we can store nanosecond pulses in a waveguide with continuous tunable storage time while maintaining the pulse shape. In this section we show that we can extend the operational bandwidth of our memory much further, allowing the storage of sub-ns pulses with different amplitude levels. Furthermore we show that the transfer process of photon to phonon back to photon is fully coherent, enabling the storage of different phase states. These breakthrough demonstrations show a significant increase in the capacity of the memory. The retrieval of the amplitude and phase information of two short optical pulses with 500\,ps pulse width after 3.5\,ns is shown in figure \ref{result2}a) and c). For the storage of these short pulses we used 24\,cm long waveguides, hence a better signal-to-noise ratio (due to lower overall propagation loss) is achieved in comparison to the measurements presented in figure \ref{result}. The pulse width corresponds to a bandwidth of more than 1.5\,GHz, almost 2-orders of magnitude wider than the intrinsic Brillouin linewidth. This implies a very high local Brillouin gain in the pulse overlap region as the Brillouin gain is spread out over a wide frequency range. \\
We encoded 6 different amplitude levels in pulse 1, while maintaining the amplitude level of a 2nd data pulse constant as a reference. A comparison of the original and retrieved pulse 1 (inset) shows that we can easily distinguish 6 different amplitude levels; this can be enhanced with a more sensitive detection system. The amplitude of the second retrieved data pulse remains constant as does its original amplitude. We simulate our system using coupled-mode equations\cite{Winful2013a,Winful2015} and see great agreement with our measurements, presented in figure \ref{result2}b) (more details on the simulation methods can be found in the supplementary material). \\
Further to multiple amplitude levels, we can also store and retrieve different optical phases to show the coherence of the state transfer between traveling acoustic and optical waves. To distinguish the phase we replace the direct detection scheme (single photodiode) with an interferometric homodyne detection scheme. Here, the phase encoded signal interferes with a local oscillator and is detected by a balanced detector measuring the difference signal of two equal photodiodes.

Two pulses are encoded with two different phases, either 0 and $\pi$ (blue) or $\pi$ and 0 (red), respectively figure \ref{result2}c). After being stored for 3.5\,ns, these same values can be read out (light blue and orange) and are clearly distinguishable. For a better understanding, the states in the phase space (I - Q diagram) are related to the optical pulses. For phase 0, the local oscillator and the data pulses interfere constructively, resulting in a positive value, for $\pi$ they interfere destructively which results in a negative pulse on the balanced detector. The phase retrieval is possible due to the coherence of the Brillouin process and proves its potential as a coherent buffer. Note, that this feature can be implemented for any phase in the entire phase space and not only for 0 and $\pi$. Here, too, as for the amplitude measurements, we simulate our system and see excellent agreement between the measurements and the simulations (figure \ref{result2}d). \newline \indent

\subsection{Multi-wavelength operation}

\begin{figure*}[!ht]
  \centering
  \includegraphics[width=0.99\textwidth]{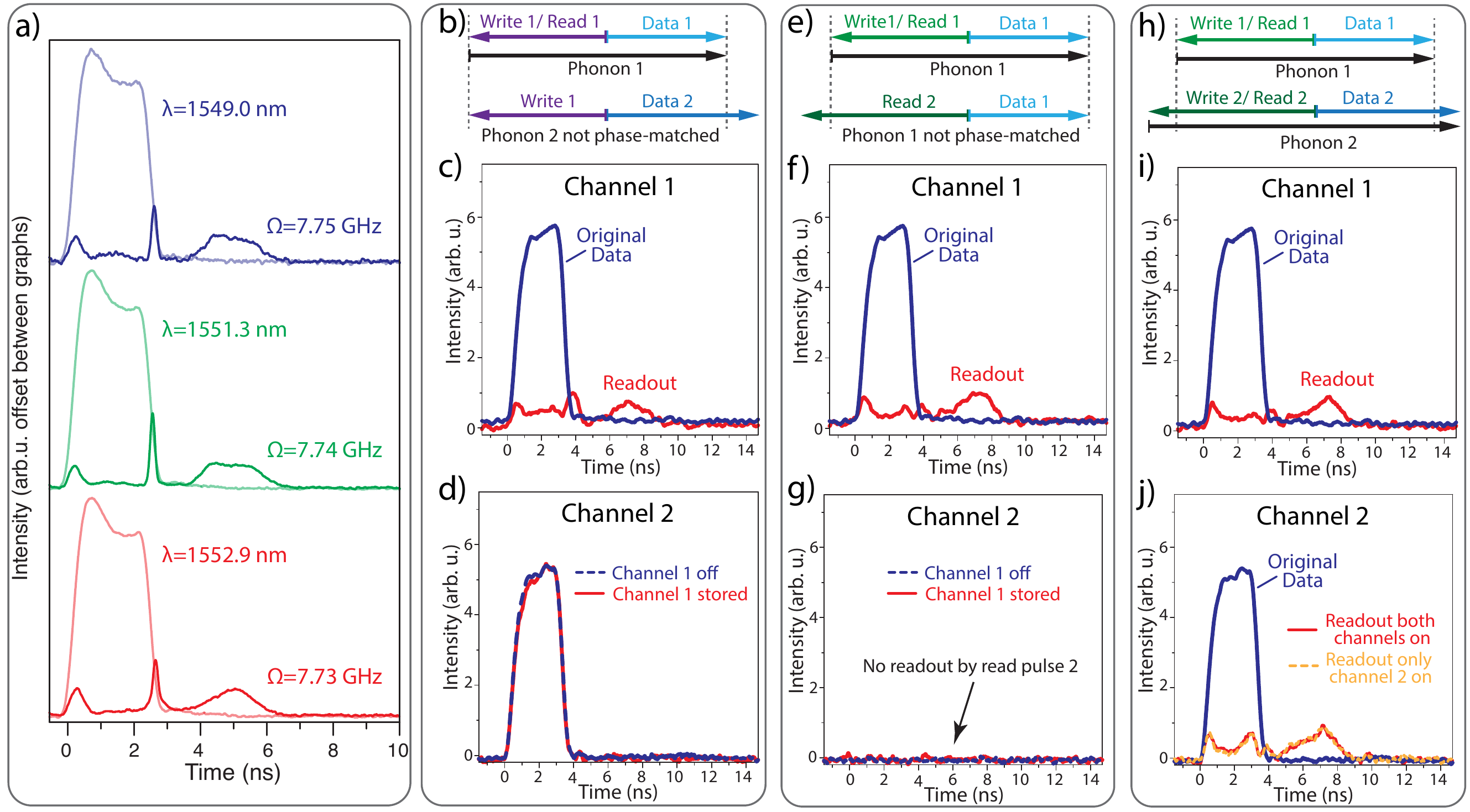}
\caption{\textbf{Multi wavelength operation} \small{a) Multi wavelength operation of the photonic-phononic memory for three different laser wavelengths. The efficiency of the memory remains the same. b) Phase matching condition for two data pulses and one write pulse phase-matched to data pulse 1 (corresponds to measurement c) and d)). The second data pulse is not phase-matched and therefore not affected by the write pulse. c) Storage and retrieval of data pulse channel 1 while a second data pulse (d) in a separate channel (100\,GHz away from channel 1) is unaffected. e) Phase matching condition for two separate read pulses with a data pulse and a phonon (corresponds to measurement f) and g)). One can see that the second read pulse cannot readout the phonon. f) Shows the writing and retrieving of a data pulse in channel 1 while at the same time no readout pulse can be seen in channel 2 (g). h) Phase matching condition for two channels operating at two different wavelengths. i) Storage and retrieval of data in channel 1 while simultaneously storing and retrieving data in channel 2 (j). Note: The difference in the noise floor between channel\,1 and channel\,2 is caused by the different noise properties of the two different photodiodes.}}
\label{multi}
\end{figure*}

Here we demonstrate the multi-wavelength capabilities of our memory. On the one hand the memory operation must work at several different wavelengths, while on the other hand the cross-talk between wavelength channels should be minimal. Our Brillouin-based memory works at all wavelengths where the waveguide is transparent, in contrast to resonator-based approaches where one is bound to the particular resonance frequencies. This transparency window reaches in the case of chalcogenide from the visible all the way to the deep infrared. To demonstrate the wavelength multiplexing capacity, we adjusted the operation laser wavelength to 3 different values in the tuning range of our laser. It can be seen from figure \ref{multi}a) that the same efficiency is achieved for all wavelengths. The pulse shape for the 1552.9\,nm measurement is slightly distorted which can be assigned to the limitations in our setup (power limitations and effects of the nonlinear loop (see supplementary)) and is not of fundamental nature. Every pair of frequencies (data frequency and read/write frequency) excites an acoustic wave at a specific frequency, which can be most easily seen in the equation for the Brillouin frequency shift $\Omega= 2 V_{A} n_{\mathrm{eff}} / \lambda$  ($\Omega$ Brillouin frequency shift, $V_{A}$ longitudinal acoustic velocity, $n_{\mathrm{eff}}$ effective refractive index and $\lambda$ laser wavelength). The respective Brillouin shifts $\Omega$ are indicated in figure \ref{multi}a).\\
The second important point concerns the cross-talk between different wavelength channels: Here, the unique phase matching conditions between traveling acoustic and optical waves inhibits mixing of different frequency channels (as illustrated in figure \ref{multi} b), e) and h)). The storage and read-out process is strictly bound to specific phase matching conditions, such that the process is operational at different wavelengths at the same time. To prove the point that there is no cross-talk between different channels we couple two data pulses, separated by only 100\,GHz, simultaneously into the waveguide and measured the waveguide output using a dual channel oscilloscope (figure~\ref{multi} c) and d)). When adding write and read pulses phase-matched to the data pulses in channel 1 only the data pulse in this channel gets stored and retrieved (figure~\ref{multi} c)) while there is no effect observable in channel 2 (figure~\ref{multi}d)). This result shows that one can operate the memory on individual data streams, separated by a standard 100\,GHz guard-band, without adding any detrimental distortions on the other channel. \\
We furthermore experimentally show that a non-phase-matched read pulse cannot retrieve information stored in a different frequency channel (figure~\ref{multi} f) and g)). To demonstrate this, we store and retrieve an optical data pulse in channel 1 (figure \ref{multi} f), while simultaneously a second read pulse, separated by 100\,GHz from the read pulse in channel 1, does not readout the stored data pulse, see figure \ref{multi} g). This is a major difference to light storage schemes based on opto-mechanical resonator scheme where light interacts with standing acoustic waves or couples to transverse acoustic modes, as in this cases there is no or only a minimum momentum transfer. Therefore many different optical modes get modulated by the presence of the acoustic mode, hence these schemes are predestined for wavelength conversion~\cite{Fan2016}. \\
Finally, we show that there is no cross-talk between the two wavelength channels separated by 100\,GHz even when optical data pulses are stored and retrieved simultaneously in the two channels (figure \ref{multi} i) and j)). For comparison figure \ref{multi} j) shows also the stored and retrieved data pulse with the second channel turned off (orange dashed line).

\subsection{Discussion}

In this article we have demonstrated a coherent photonic memory based on optically actuated traveling acoustic phonons in a planar waveguide. Our memory relies on a state transfer from photons to slowly propagating phonons and can therefore be seen as a completely different approach compared to schemes which rely on a reduced group velocity of light pulses such as coupled optical resonators\cite{Lee2006,Cardenas2010,Xia2007}, photonic crystal cavities\cite{Kuramochi2014,Baba2008,Krauss2007} or slow-light schemes\cite{Thevenaz2008,Baba2008,Okawachi2005,Gersen2005}. By transferring the optical pulse to an acoustic wave, our optical buffer allows to circumvent detrimental optical dispersion effects\cite{Khurgin2006,Khurgin2005} and allows for relatively long delay times of many pulse widths. The delay time can potentially be increased even further using a cascaded process\cite{Fan2015} or by further engineering the dissipation rate of the traveling acoustic phonon.\\
The photonic-phononic memory is fully controlled by the spatial-temporal overlap of the data, write and read optical pulses in a simple planar photonic circuit. Therefore the buffer is not an additional element of the circuit, but the photonic waveguide/link itself can be used as the buffering element bringing additional functionality to optical interconnects for next generation microelectronic networks\cite{Miller2010a,Alduino2007,Lee2008b,Miller2009}.\\
Storing and retrieving the full coherent information carried by the light signal enables the processing of multiple amplitude and phase levels, which is essential for contemporary communications schemes and greatly increases the number of bits that can be stored. The ultrahigh Brillouin gain in chalcogenide glass allows for the encoding of signals down to 500\,ps pulse width. Even shorter pulses can be realized by further reducing waveguide losses and increasing the opto-acoustic coupling by tailoring acoustic properties. A very important feature of the the demonstrated buffer is the operation at separate optical wavelengths without crosstalk. In particularly the frequency preserving property due to the stringent phase-matching condition between traveling acoustic and optical waves is essential for multi-channel operation in order to store and retrieve information at different frequency channels unambiguously. In communication networks and computing architectures, this versatility plus the continuous tunability of the storage time of up to several nanoseconds enables precise and dynamic synchronization of optical data streams between several high-speed parallel processes.

\section*{methods}
\label{methods}
\subsection{Experimental set-up for light storage}

A narrow-linewidth distributed feedback (DFB) laser at 1550\,nm is divided into two arms - data and write\,$\&$\,read arm - where the data pulse is frequency up-shifted by the Brillouin frequency shift $\Omega$ via a single-sideband modulator. The pulses are imprinted by two intensity modulators connected to a short-pulse generator. The write\,$\&$\,read pulses are amplified by an Erbium doped fiber amplifier (EDFA). The amplified write and read pulses pass through a nonlinear fiber loop. The loop has two effects: firstly, it allows only the pulses to be transmitted and efficiently suppresses any noise or coherent background present from the laser or amplifier, respectively. Secondly, it improves the pulse shape by smoothing the edges of the pulses. After the loop a second EDFA amplifies the pulses again to reach the necessary peak power of several Watts. Bandpass filters (bandwidth 0.5\,nm) are used in both arms to minimize the white noise from the EDFAs. Both paths lead to opposite sides of the photonic chip. The original and retrieved data pulses are observed by a 12\,GHz photodiode connected to the oscilloscope. Before the photodiode a tunable narrowband filter is used to assure that only the data pulses reach the photodetector.

\subsection{Detection scheme for phase encoded signals}
For the detection of different optical phases we use a homodyne detection scheme. A local oscillator (continuous wave) at the wavelength of the data pulses interferes at a 50:50 coupler with the original and retrieved data pulses. The beat signal is sent to a polarization beam splitter both output signals of which are connected to a balanced photodetector. The polarization of the local oscillator and the data pulses are controlled such that the difference signal of both photodiodes of the balanced photodetector is maximized in order to distinguish the two phases, shifted by $\pi$.

\begin{acknowledgments} 
This work was sponsored by the Australian Research Council (ARC) Laureate Fellowship (FL120100029) and the Centre of Excellence program (CUDOS CE110001010).
\end{acknowledgments}

\section*{Author contributions}
M.M. and B.S. contributed equally to this work. M.M., B.S. and B.J.E. discussed the general idea and conceived the experiment. M.M. and B.S. designed the experiment and did the measurements and data analysis. K.V. and S.J.M. developed and fabricated the photonic chip. B.J.E. and B.S. supervised the project. All authors discussed the results. B.S., M.M. and B.J.E. wrote the manuscript with input from all co-authors.

\section*{Additional information}
Correspondence and requests for materials should be addressed to B.S. and M.M..~(email: birgit.stiller@sydney.edu.au, moritz.merklein@sydney.edu.au).

\onecolumngrid
\appendix

\section{Experimental methods}
\subsection{Experimental set-up}
\begin{figure}
  \centering
  \includegraphics[width=1\textwidth]{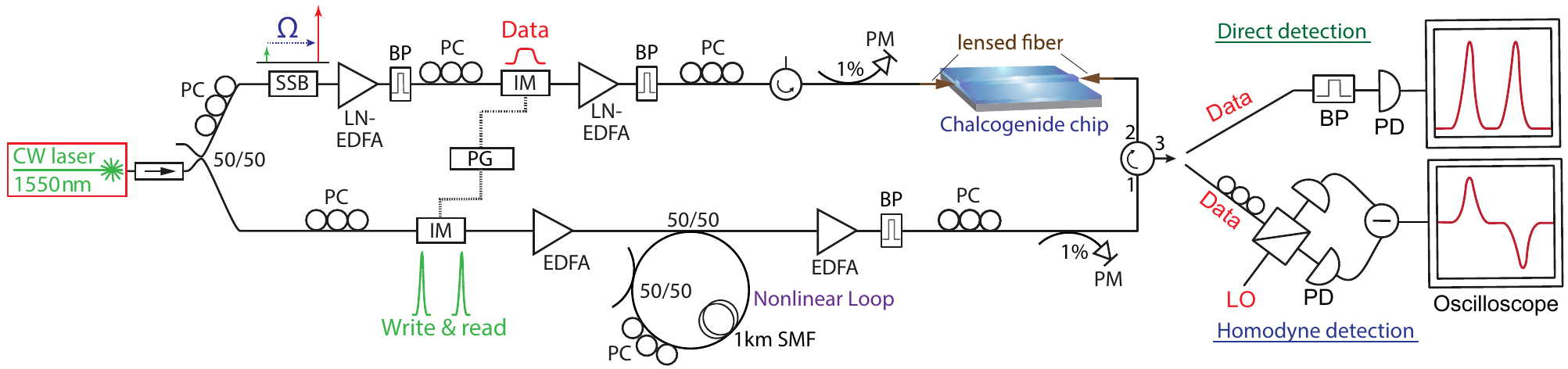}{}
\caption{\textbf{Experimental set-up} \small{CW laser: continuous wave laser; PC: polarization controller; SSB: single sideband modulator; EDFA: erbium doped fiber amplifier;  LN-EDFA: low-noise EDFA; BP: bandpass filter; SMF: standard single mode fiber; PM: power meter; LO: local oscillator; PD: photodetector}}
\label{set}
\end{figure}

The experimental set-up is shown in figure \ref{set}. A continuous wave (CW) narrow-linewidth distributed feedback (DFB) laser at 1550\,nm is divided into a data and write\,$\&$\,read arm by a 50/50 fiber coupler. The data pulses are frequency up-shifted by the Brillouin frequency shift $\Omega = 7.7$~GHz via a single-sideband modulator. The CW laser signal is carved into pulses by two intensity modulators connected to a short-pulse generator, allowing the generation of pulses with different amplitude levels and phase states. The pulses are amplified using Erbium doped fiber amplifiers (EDFA) and sub-sequentially filtered by narrow bandwidth (0.5\,nm) bandpass filters to reduce the effect of broadband white noise introduced by the amplification step. Additionally to the passive bandpass filter a nonlinear fiber loop is implemented in the write and read arm. The loop consists of 1\,km standard single mode fiber, a polarization controller and a 50/50 coupler to introduce some asymmetry in the two paths. This fiber loop is used for two reasons: firstly it allows only the pulses to be transmitted and efficiently suppresses any noise or coherent background present from the laser or amplifier respectively. Secondly, it improves the pulse shape by smoothing the edges of the pulses. After the loop a second EDFA amplifies the pulses again to reach the necessary peak power of several Watts. Both paths lead to opposite sides of the photonic chip and are coupled to the waveguide using lensed fibers.\\
For the multi-wavelenght measurement a second laser, 100\,GHz apart from the first laser, is used and coupled into the data or write and read arm, respectively. The output from the chip (circulator port 3 in figure \ref{set}) is split with a 50\,/\,50 fiber coupler and send to two narrowband filters to separate the two wavelength channels. Each filtered channel is then detected using two 12\,GHz photodiodes connected to a dual channel oscilloscope.

\subsection{Detection schemes}
Two different detection schemes are used to detect the transmitted and retrieved data pulses: direct detection with a single photodiode is used for the amplitude retrieval whereas a homodyne detection scheme is used for the phase measurements. For the direct detection scheme a 12\,GHz photodiode connected to the oscilloscope is used. For the homodyne detection scheme a local oscillator (continuous wave) at the wavelength of the data pulses interferes at a 50:50 coupler with the original and retrieved data pulses. The beat signal is sent to a polarization beam splitter both output signals of which are connected to a balanced photodetector. The polarization of the local oscillator and the data pulses are controlled such that the difference signal of both photodiodes of the balanced photodetector is maximized in order to distinguish the two phases, shifted by $\pi$. In both detection schemes a tunable narrow-band filter ($\approx 4$~GHz) is used to assure that only the data pulses reach the photodetector.

\section{Additional measurements}
\subsection{Acoustic decay time:} 
\begin{figure}[h!]
  \centering
  \includegraphics[width=0.75\textwidth]{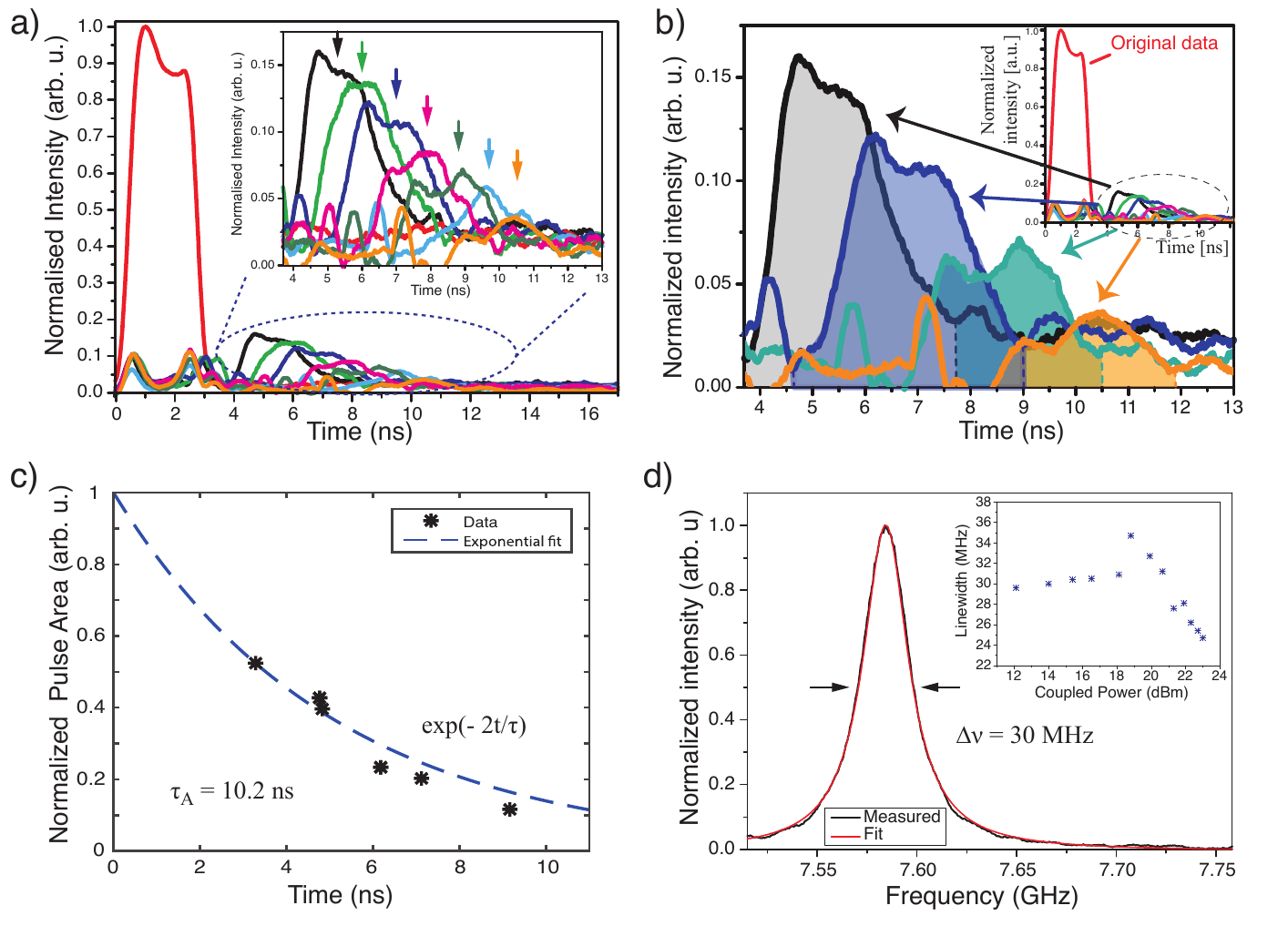}
\caption{\textbf{Storage time and acoustic decay time} \small{a) Measurements of different storage times. b) Four retrieved data pulses with shaded pulse area. c) Pulse area with exponential fit reveals acoustic decay time of 10.2\,ns. d) Brillouin linewidth measured with a CW pump-probe set-up showing 30\,MHz linewidth. The inset shows the linewidth for different pump powers.}}
\label{dec}
\end{figure}

We analyzed the storage time of the phonon memory and compared it with a standard pump probe measurement of the Brillouin linewidth. Figure \ref{dec}a) shows different readout pulses for different storage times up to 10.5\,ns. For increasing storage times the readout efficiency decreases due to the decaying amplitude of the acoustic wave. The area of the retrieved pulses is integrated to determine the acoustic decay time (figure \ref{dec}b). The exponential decrease of the pulse areas are plotted in figure \ref{dec}c) and an exponential fit $exp(-2 t/\tau_\mathrm{A})$ reveals an acoustic decay time of 10.2\,ns.
We confirm the acoustic decay time by measuring the linewidth of the Brillouin gain response (figure \ref{dec}c). A modified version of the set-up in figure \ref{set} is used to execute the CW pump probe measurements. The intensity modulators were removed and the SSB modulator is frequency swept by the RF output of a vector network analyzer (VNA) to generate a seed signal. The transmitted seed signal is detected by a photodiode connected to the VNA and the Brillouin response for different pump powers is measured. The Brillouin gain linewidth $\nu_\mathrm{B}$ relates to the acoustic decay time $\tau_\mathrm{A}$ as $\tau_\mathrm{A} = 1/ \pi \Delta \nu_\mathrm{B}$. The fit in figure \ref{dec}c) shows a linewidth of 30\,MHz for 21\,dBm pump power, which agrees perfectly with the 10.2\,ns decay time measured in the storage experiment. As expected the linewidth of the gain peak decreases above the Brillouin threshold (inset figure \ref{dec}d). \\
The great agreement of the two measurement techniques not only confirms the consistency of the light storage measurements, but also suggests itself for using the storage technique to \textit{locally} access material and structure specific acoustic decay times. Whereas CW based pump probe schemes only provide an average of the acoustic decay time over the whole length of the waveguide, the pulsed measurement determines the acoustic decay time at the point of the waveguide where the pulses overlap. The overlap can be scanned along the waveguide providing spatial information about the waveguide. \\
\newpage
\subsection{Maximum amplitude readout} 
\begin{figure}[h]
  \centering
  \includegraphics[width=0.35\textwidth]{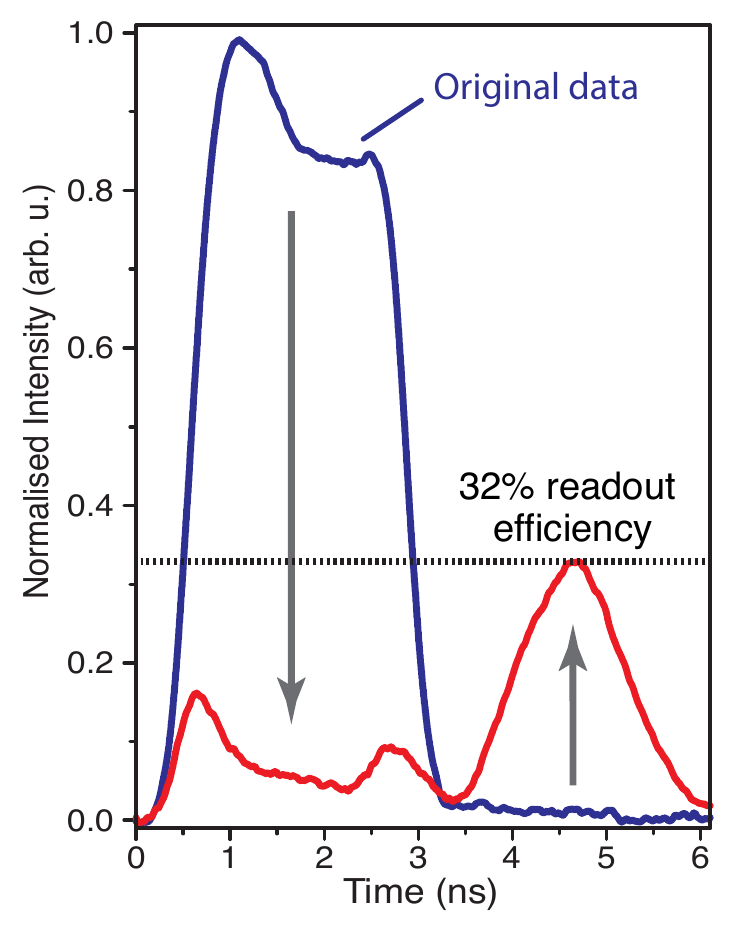}
\caption{\textbf{Maximising the amplitude retrieval efficiency} \small{A 32\% retrieval efficiency of the pulse amplitude after a storage time of 3.5\,ns was achieved}}
\label{rec}
\end{figure}

The readout efficiency could be increased to 32\% after 3.5\,ns storage time as shown in figure \ref{rec}. It is known from numerical studies that the amplitude of the retrieved pulses can be enhanced by using chirped pulses \cite{Winful2013a}. This can greatly enhance the performance of the phononic buffer for simple on-off data streams, where one is only interested in retrieving the amplitude. In this case it also could increase the maximum retrieval time of the buffer by lifting the retrieved amplitude above the noise floor. The 32\% readout depicted in figure \ref{rec} is achieved by increasing the input power into the nonlinear loop (see set-up figure \ref{set}). Higher input power increases the nonlinear process known as self phase modulation, chirping the pulses \cite{Boyd2003}. The nonlinear loop in the set-up therefore not only reduces the noise, but also allows for a more efficient readout amplitude through compression of the retrieved pulse. However the pulse shape is not maintained in this case.
Besides using chirped pulses to improve the maximum retrieval amplitude, it was also shown theoretically that small amounts of chirp help to more efficiently excite the acoustic wave  \cite{Winful2015}. This can be understood by drawing an analogy to the McCall and Hahn area theorem for atomic two-level system  \cite{McCall1969}. Analogues to the $\pi$ pulse in atomic resonances, a normalized "pulse area" of the write pulses can be defined and is given by \cite{Winful2015} $\Theta = \sqrt{g_\mathrm{B} c / 8 A_\mathrm{eff} \tau_\mathrm{B} n} \times \int A(t) dt$, with the Brillouin gain coefficient $g_\mathrm{B}$, the speed of light $c$, the effective mode area $A_\mathrm{eff}$, the acoustic decay time $\tau_\mathrm{B}$, the refractive index  $n$ and the time integral over the pulse envelope $A(t)$. The maximum efficiency for exciting the acoustic wave is achieved when $\Theta = (m + 1/2)\pi$ with $m$ being an integer number. However the data pulse cannot be transferred to the acoustic wave if the "pulse area" is an integer multiple of $\pi$. If the pulse area is a multiple of $\pi$ the first half of the pulse will write the acoustic wave, while the second half retrieves it again. However, for a linear chirped pulse the beginning of the pulse has a different frequency as the end of the pulse. Therefore, only a certain part of the pulse resonantly excites the acoustic wave and importantly does not de-excite the acoustic wave.

\section{Simulation method}

To simulate the phase and amplitude response of our system we solved standard coupled mode equations as presented in Ref. [1] using an implicit fourth order Runge-Kutta method\cite{Sterke1991}. The slowly-varying envelope coupled mode equations for a forward traveling pump wave $A_{\mathrm{P}}$, a counterpropagating Stokes wave $A_{\mathrm{S}}$ and an acoustic wave Q can be written in the following form\cite{Winful2013a}:
\begin{equation}
\frac{\partial A_{\mathrm{P}}}{\partial z} + \frac{n}{c} \frac{\partial A_{\mathrm{P}}}{\partial t}  = -\frac{g_{\mathrm{0}}}{2 A_{\mathrm{eff}}} Q A_{\mathrm{S}} - \frac{1}{2} \alpha A_{\mathrm{P}}
\end{equation}
\begin{equation}
- \frac{\partial A_{\mathrm{S}}}{\partial z} + \frac{n}{c} \frac{\partial A_{\mathrm{S}}}{\partial t}  = \frac{g_{\mathrm{0}}}{2 A_{\mathrm{eff}}} Q^{*} A_{\mathrm{P}} - \frac{1}{2} \alpha A_{\mathrm{S}}
\end{equation}
\begin{equation}
2 \tau_{\mathrm{B}} \frac{\partial Q}{\partial t} + Q = A_{\mathrm{P}} A_{\mathrm{P}}^{*}
\end{equation}

The slowly varying envelopes $A_{\mathrm{P}}$, $A_{\mathrm{S}}$ are normalized such that $|A_{\mathrm{P/S}}|^{2}$ is the power in Watts, Q is the amplitude of the acoustic wave, n is the refractive index, c the speed of light, $g_{\mathrm{0}}$ the Brillouin gain coefficient, $A_{\mathrm{eff}}$ the effective mode area, $\tau_{\mathrm{B}}$ the acoustic lifetime and $\alpha$ the waveguide loss parameter.

The envelopes of the input data, write and read pulses are approximated to have Gaussian form\cite{Winful2015}:
\begin{equation}
A_{\mathrm{P/S}} = A_{0} \mathrm{exp} (- \frac{1+i C}{2} \frac{t^{2}}{\tau^{2}})
\end{equation}

with the parameter C giving the chirp rate in GHz\,/\,ns following the definition of Ref.\cite{Winful2015} and $\tau$ being the FWHM. The parameters used for the amplitude simulations (Fig. 2b) are as follows: n = 2.4, $g_{\mathrm{0}} = 0.715 \cdot 10^{-9} \mathrm{m/W}$, $A_{\mathrm{eff}} = 1.5 \cdot 10^{-15} \mathrm{m^{2}}$, $\tau_{\mathrm{B}} = 10.5\, \mathrm{ns}$, $\alpha = \mathrm{0.2\,dB/cm}$. The FWHM of the data pulses is 500\,ps and the peak power is varying in equidistant steps from 15\,mW to 40\,mW. The FWHM of the data pulses was 1\,ns, the peak power 3.5\,W and C = 0.88\,GHz\,/\,ns. The temporal separation of the write and the read pulse was 3.5\,ns. The parameters used for the phase simulations (Fig. 2d) are the same as for the amplitude simulations with 40\,mW of data power and two different phases 0 and $\pi$.

\begin{spacing}{1.25}

\end{spacing} 
 

\begin{thebibliography}{10}
\providecommand{\url}[1]{#1}
\csname url@samestyle\endcsname
\providecommand{\newblock}{\relax}
\providecommand{\bibinfo}[2]{#2}
\providecommand{\BIBentrySTDinterwordspacing}{\spaceskip=0pt\relax}
\providecommand{\BIBentryALTinterwordstretchfactor}{4}
\providecommand{\BIBentryALTinterwordspacing}{\spaceskip=\fontdimen2\font plus
\BIBentryALTinterwordstretchfactor\fontdimen3\font minus
  \fontdimen4\font\relax}
\providecommand{\BIBforeignlanguage}[2]{{%
\expandafter\ifx\csname l@#1\endcsname\relax
\typeout{** WARNING: IEEEtran.bst: No hyphenation pattern has been}%
\typeout{** loaded for the language `#1'. Using the pattern for}%
\typeout{** the default language instead.}%
\else
\language=\csname l@#1\endcsname
\fi
#2}}
\providecommand{\BIBdecl}{\relax}
\BIBdecl

\bibitem{Weis2010}
\bibinfo{author}{Weis, S.} \emph{et~al.}
\newblock \bibinfo{title}{{Optomechanically Induced Transparency}}.
\newblock \emph{\bibinfo{journal}{Science}} \textbf{\bibinfo{volume}{330}},
  \bibinfo{pages}{1520--1523} (\bibinfo{year}{2010}).

\bibitem{Safavi-Naeini2011}
\bibinfo{author}{Safavi-Naeini, A.~H.} \emph{et~al.}
\newblock \bibinfo{title}{{Electromagnetically induced transparency and slow
  light with optomechanics.}}
\newblock \emph{\bibinfo{journal}{Nature}} \textbf{\bibinfo{volume}{472}},
  \bibinfo{pages}{69--73} (\bibinfo{year}{2011}).

\bibitem{Verhagen2012}
\bibinfo{author}{Verhagen, E.}, \bibinfo{author}{Del{\'{e}}glise, S.},
  \bibinfo{author}{Weis, S.}, \bibinfo{author}{Schliesser, a.} \&
  \bibinfo{author}{Kippenberg, T.~J.}
\newblock \bibinfo{title}{{Quantum-coherent coupling of a mechanical oscillator
  to an optical cavity mode.}}
\newblock \emph{\bibinfo{journal}{Nature}} \textbf{\bibinfo{volume}{482}},
  \bibinfo{pages}{63--7} (\bibinfo{year}{2012}).

\bibitem{Fiore2011a}
\bibinfo{author}{Fiore, V.} \emph{et~al.}
\newblock \bibinfo{title}{{Storing optical information as a mechanical
  excitation in a silica optomechanical resonator}}.
\newblock \emph{\bibinfo{journal}{Physical Review Letters}}
  \textbf{\bibinfo{volume}{107}}, \bibinfo{pages}{1--5} (\bibinfo{year}{2011}).

\bibitem{Galland2014}
\bibinfo{author}{Galland, C.}, \bibinfo{author}{Sangouard, N.},
  \bibinfo{author}{Piro, N.}, \bibinfo{author}{Gisin, N.} \&
  \bibinfo{author}{Kippenberg, T.~J.}
\newblock \bibinfo{title}{{Heralded single-phonon preparation, storage, and
  readout in cavity optomechanics}}.
\newblock \emph{\bibinfo{journal}{Physical Review Letters}}
  \textbf{\bibinfo{volume}{112}}, \bibinfo{pages}{1--6} (\bibinfo{year}{2014}).

\bibitem{Wang2013b}
\bibinfo{author}{Dong, C.}, \bibinfo{author}{Fiore, V.},
  \bibinfo{author}{Kuzyk, M.~C.} \& \bibinfo{author}{Wang, H.}
\newblock \bibinfo{title}{{Optomechanical Dark Mode}}.
\newblock \emph{\bibinfo{journal}{Science}} \textbf{\bibinfo{volume}{338}},
  \bibinfo{pages}{1609--1613} (\bibinfo{year}{2012}).

\bibitem{Dong2015}
\bibinfo{author}{Dong, C.-H.} \emph{et~al.}
\newblock \bibinfo{title}{{Brillouin-scattering-induced transparency and
  non-reciprocal light storage}}.
\newblock \emph{\bibinfo{journal}{Nature Communications}}
  \textbf{\bibinfo{volume}{6}} (\bibinfo{year}{2015}).

\bibitem{Kim2015}
\bibinfo{author}{Kim, J.}, \bibinfo{author}{Kuzyk, M.~C.},
  \bibinfo{author}{Han, K.}, \bibinfo{author}{Wang, H.} \&
  \bibinfo{author}{Bahl, G.}
\newblock \bibinfo{title}{{Non-reciprocal Brillouin scattering induced
  transparency}}.
\newblock \emph{\bibinfo{journal}{Nature Physics}} \bibinfo{pages}{1--6}
  (\bibinfo{year}{2015}).

\bibitem{Shen2016}
\bibinfo{author}{Shen, Z.} \emph{et~al.}
\newblock \bibinfo{title}{{Experimental realization of optomechanically induced
  non-reciprocity}}.
\newblock \emph{\bibinfo{journal}{Nature Photonics}}
  \textbf{\bibinfo{volume}{10}}, \bibinfo{pages}{1--5} (\bibinfo{year}{2016}).

\bibitem{Fang2016a}
\bibinfo{author}{Fang, K.}, \bibinfo{author}{Matheny, M.~H.},
  \bibinfo{author}{Luan, X.} \& \bibinfo{author}{Painter, O.}
\newblock \bibinfo{title}{{Optical transduction and routing of microwave
  phonons in cavity-optomechanical circuits}}.
\newblock \emph{\bibinfo{journal}{Nature Photonics}}
  \textbf{\bibinfo{volume}{10}}, \bibinfo{pages}{489--496}
  (\bibinfo{year}{2016}).

\bibitem{Balram2015}
\bibinfo{author}{Balram, K.~C.}, \bibinfo{author}{Davan{\c{c}}o, M.~I.},
  \bibinfo{author}{Song, J.~D.} \& \bibinfo{author}{Srinivasan, K.}
\newblock \bibinfo{title}{{Coherent coupling between radiofrequency, optical
  and acoustic waves in piezo-optomechanical circuits}}.
\newblock \emph{\bibinfo{journal}{Nature Photonics}}
  \textbf{\bibinfo{volume}{10}}, \bibinfo{pages}{346--352}
  (\bibinfo{year}{2016}).

\bibitem{Li2015a}
\bibinfo{author}{Li, H.}, \bibinfo{author}{Tadesse, S.~A.},
  \bibinfo{author}{Liu, Q.} \& \bibinfo{author}{Li, M.}
\newblock \bibinfo{title}{{Nanophotonic cavity optomechanics with propagating
  acoustic waves at frequencies up to 12 GHz}}.
\newblock \emph{\bibinfo{journal}{Optica}} \textbf{\bibinfo{volume}{2}},
  \bibinfo{pages}{826} (\bibinfo{year}{2015}).

\bibitem{Fiore2013}
\bibinfo{author}{Fiore, V.}, \bibinfo{author}{Dong, C.},
  \bibinfo{author}{Kuzyk, M.~C.} \& \bibinfo{author}{Wang, H.}
\newblock \bibinfo{title}{{Optomechanical light storage in a silica
  microresonator}}.
\newblock \emph{\bibinfo{journal}{Physical Review A}} \textbf{\bibinfo{volume}{87}}, \bibinfo{pages}{1--6}
  (\bibinfo{year}{2013}).

\bibitem{Safavi-Naeini2011b}
\bibinfo{author}{Safavi-Naeini, A.~H.} \& \bibinfo{author}{Painter, O.}
\newblock \bibinfo{title}{{Proposal for an optomechanical traveling wave
  phonon-photon translator}}.
\newblock \emph{\bibinfo{journal}{New Journal of Physics}}
  \textbf{\bibinfo{volume}{13}} (\bibinfo{year}{2011}).

\bibitem{Chang2010a}
\bibinfo{author}{Chang, D.~E.}, \bibinfo{author}{Safavi-Naeini, A.~H.},
  \bibinfo{author}{Hafezi, M.} \& \bibinfo{author}{Painter, O.}
\newblock \bibinfo{title}{{Slowing and stopping light using an optomechanical
  crystal array}}.
\newblock \emph{\bibinfo{journal}{New Journal of Physics}}
  \textbf{\bibinfo{volume}{13}}, \bibinfo{pages}{023003}
  (\bibinfo{year}{2011}).

\bibitem{Hill2012}
\bibinfo{author}{Hill, J.~T.}, \bibinfo{author}{Safavi-Naeini, A.~H.},
  \bibinfo{author}{Chan, J.} \& \bibinfo{author}{Painter, O.}
\newblock \bibinfo{title}{{Coherent optical wavelength conversion via cavity
  optomechanics.}}
\newblock \emph{\bibinfo{journal}{Nature Communications}}
  \textbf{\bibinfo{volume}{3}}, \bibinfo{pages}{1196} (\bibinfo{year}{2012}).

\bibitem{Fan2016}
\bibinfo{author}{Fan, L.} \emph{et~al.}
\newblock \bibinfo{title}{{Integrated optomechanical single-photon frequency shifter}}. 
\newblock \emph{\bibinfo{journal}{Nature Photonics}}
  \textbf{\bibinfo{volume}{10}}, \bibinfo{pages}{766--770}
  (\bibinfo{year}{2016}).

\bibitem{MingCaiOskarPainter2000}
\bibinfo{author}{Cai, M.}, \bibinfo{author}{Painter, O.} \&
  \bibinfo{author}{Vahala, K.~J.}
\newblock \bibinfo{title}{{Observation of Critical Coupling in a Fiber Taper to
  a Silica-Microsphere Whispering-Gallery Mode System}}.
\newblock \emph{\bibinfo{journal}{Physical Review Letters}}
  \textbf{\bibinfo{volume}{85}}, \bibinfo{pages}{74--77}
  (\bibinfo{year}{2000}).

\bibitem{Vahala2003}
\bibinfo{author}{Vahala, K.~J.}
\newblock \bibinfo{title}{{Optical microcavities.}}
\newblock \emph{\bibinfo{journal}{Nature}} \textbf{\bibinfo{volume}{424}},
  \bibinfo{pages}{839--46} (\bibinfo{year}{2003}).

\bibitem{Safavi-Naeini2010}
\bibinfo{author}{Safavi-Naeini, A.~H.}, \bibinfo{author}{Alegre, T. P.~M.},
  \bibinfo{author}{Winger, M.} \& \bibinfo{author}{Painter, O.}
\newblock \bibinfo{title}{{Optomechanics in an ultrahigh- Q two-dimensional
  photonic crystal cavity}}.
\newblock \emph{\bibinfo{journal}{Applied Physics Letters}}
  \textbf{\bibinfo{volume}{97}}, \bibinfo{pages}{97--99}
  (\bibinfo{year}{2010}).

\bibitem{Boyd2003}
\bibinfo{author}{Boyd, R.~W.}
\newblock \emph{\bibinfo{title}{{Nonlinear Optics}}} (\bibinfo{publisher}{Acad.
  Press}, \bibinfo{year}{2003}).

\bibitem{Zhu2007}
\bibinfo{author}{Zhu, Z.}, \bibinfo{author}{Gauthier, D.~J.} \&
  \bibinfo{author}{Boyd, R.~W.}
\newblock \bibinfo{title}{{Stored light in an optical fiber via stimulated
  Brillouin scattering.}}
\newblock \emph{\bibinfo{journal}{Science}}
  \textbf{\bibinfo{volume}{318}}, \bibinfo{pages}{1748--50}
  (\bibinfo{year}{2007}).

\bibitem{Pant2011}
\bibinfo{author}{Pant, R.} \emph{et~al.}
\newblock \bibinfo{title}{{On-chip stimulated Brillouin scattering}}.
\newblock \emph{\bibinfo{journal}{Optics Express}}
  \textbf{\bibinfo{volume}{19}}, \bibinfo{pages}{8285--8290}
  (\bibinfo{year}{2011}).

\bibitem{VanLaer2015}
\bibinfo{author}{{Van Laer}, R.}, \bibinfo{author}{Kuyken, B.},
  \bibinfo{author}{{Van Thourhout}, D.} \& \bibinfo{author}{Baets, R.}
\newblock \bibinfo{title}{{Interaction between light and highly confined
  hypersound in a silicon photonic nanowire}}.
\newblock \emph{\bibinfo{journal}{Nature Photonics}}
  \textbf{\bibinfo{volume}{9}}, \bibinfo{pages}{199--203}
  (\bibinfo{year}{2015}).

\bibitem{Kittlaus2015}
\bibinfo{author}{Kittlaus, E.~A.}, \bibinfo{author}{Shin, H.} \&
  \bibinfo{author}{Rakich, P.~T.}
\newblock \bibinfo{title}{{Large Brillouin amplification in silicon}}.
\newblock \emph{\bibinfo{journal}{Nature Photonics}}
  \textbf{\bibinfo{volume}{10}}, \bibinfo{pages}{463--467}
  (\bibinfo{year}{2016}).

\bibitem{Merklein2016a}
\bibinfo{author}{Merklein, M.} \emph{et~al.}
\newblock \bibinfo{title}{{Stimulated Brillouin scattering in photonic
  integrated circuits: novel applications and devices}}.
\newblock \emph{\bibinfo{journal}{IEEE Journal of Selected Topics in Quantum
  Electronics}} \textbf{\bibinfo{volume}{22}}
  (\bibinfo{year}{2016}).

\bibitem{Eggleton2013}
\bibinfo{author}{Eggleton, B.~J.}, \bibinfo{author}{Poulton, C.~G.} \&
  \bibinfo{author}{Pant, R.}
\newblock \bibinfo{title}{{Inducing and harnessing stimulated Brillouin
  scattering in photonic integrated circuits}}.
\newblock \emph{\bibinfo{journal}{Advances in Optics and Photonics}}
  \textbf{\bibinfo{volume}{5}}, \bibinfo{pages}{536--587}
  (\bibinfo{year}{2013}).

\bibitem{Santagiustina2013}
\bibinfo{author}{Santagiustina, M.}, \bibinfo{author}{Chin, S.},
  \bibinfo{author}{Primerov, N.}, \bibinfo{author}{Ursini, L.} \&
  \bibinfo{author}{Th{\'{e}}venaz, L.}
\newblock \bibinfo{title}{{All-optical signal processing using dynamic
  Brillouin gratings.}}
\newblock \emph{\bibinfo{journal}{Scientific Reports}}
  \textbf{\bibinfo{volume}{3}} (\bibinfo{year}{2013}).

\bibitem{Poulton2013a}
\bibinfo{author}{Poulton, C.~G.}, \bibinfo{author}{Pant, R.} \&
  \bibinfo{author}{Eggleton, B.~J.}
\newblock \bibinfo{title}{{Acoustic confinement and stimulated Brillouin
  scattering in integrated optical waveguides}}.
\newblock \emph{\bibinfo{journal}{Journal of the Optical Society of America B}}
  \textbf{\bibinfo{volume}{30}}, \bibinfo{pages}{2657--2664}
  (\bibinfo{year}{2013}).

\bibitem{Zhu2011}
\bibinfo{author}{Zhu, Y.}, \bibinfo{author}{Lee, M.}, \bibinfo{author}{Neifeld,
  M.~A.} \& \bibinfo{author}{Gauthier, D.~J.}
\newblock \bibinfo{title}{{High-fidelity, broadband
  stimulated-Brillouin-scattering-based slow light using fast noise
  modulation}}.
\newblock \emph{\bibinfo{journal}{Optics Express}}
  \textbf{\bibinfo{volume}{19}}, \bibinfo{pages}{687} (\bibinfo{year}{2011}).

\bibitem{Winful2013a}
\bibinfo{author}{Winful, H.}
\newblock \bibinfo{title}{{Chirped Brillouin dynamic gratings for storing and
  compressing light}}.
\newblock \emph{\bibinfo{journal}{Optics Express}}
  \textbf{\bibinfo{volume}{21}}, \bibinfo{pages}{10039--10047}
  (\bibinfo{year}{2013}).

\bibitem{Winful2015}
\bibinfo{author}{Dong, M.} \& \bibinfo{author}{Winful, H.~G.}
\newblock \bibinfo{title}{{Area dependence of chirped-pulse stimulated
  Brillouin scattering: implications for stored light and dynamic gratings}}.
\newblock \emph{\bibinfo{journal}{Journal of the Optical Society of America B}}
  \textbf{\bibinfo{volume}{32}}, \bibinfo{pages}{2514} (\bibinfo{year}{2015}).

\bibitem{Lee2006}
\bibinfo{author}{Lee, B.~G.}, \bibinfo{author}{Small, B.~a.},
  \bibinfo{author}{Bergman, K.}, \bibinfo{author}{Xu, Q.} \&
  \bibinfo{author}{Lipson, M.}
\newblock \bibinfo{title}{{Transmission of high-data-rate optical signals
  through a micrometer-scale silicon ring resonator}}.
\newblock \emph{\bibinfo{journal}{Optics Letters}}
  \textbf{\bibinfo{volume}{31}}, \bibinfo{pages}{2701} (\bibinfo{year}{2006}).

\bibitem{Cardenas2010}
\bibinfo{author}{Cardenas, J.} \emph{et~al.}
\newblock \bibinfo{title}{{Wide-bandwidth continuously tunable optical delay
  line using silicon microring resonators.}}
\newblock \emph{\bibinfo{journal}{Optics Express}}
  \textbf{\bibinfo{volume}{18}}, \bibinfo{pages}{26525--34}
  (\bibinfo{year}{2010}).

\bibitem{Xia2007}
\bibinfo{author}{Xia, F.}, \bibinfo{author}{Sekaric, L.} \&
  \bibinfo{author}{Vlasov, Y.}
\newblock \bibinfo{title}{{Ultracompact optical buffers on a silicon chip}}.
\newblock \emph{\bibinfo{journal}{Nature Photonics}}
  \textbf{\bibinfo{volume}{1}}, \bibinfo{pages}{65--71} (\bibinfo{year}{2007}).

\bibitem{Kuramochi2014}
\bibinfo{author}{Kuramochi, E.} \emph{et~al.}
\newblock \bibinfo{title}{{Large-scale integration of wavelength-addressable
  all-optical memories on a photonic crystal chip}}.
\newblock \emph{\bibinfo{journal}{Nature Photonics}}
  \textbf{\bibinfo{volume}{8}}, \bibinfo{pages}{474--481}
  (\bibinfo{year}{2014}).

\bibitem{Baba2008}
\bibinfo{author}{Baba, T.}
\newblock \bibinfo{title}{{Slow light in photonic crystals}}.
\newblock \emph{\bibinfo{journal}{Nature Photonics}}
  \textbf{\bibinfo{volume}{2}}, \bibinfo{pages}{465--473}
  (\bibinfo{year}{2008}).

\bibitem{Krauss2007}
\bibinfo{author}{Krauss, T.~F.}
\newblock \bibinfo{title}{{Slow light in photonic crystal waveguides}}.
\newblock \emph{\bibinfo{journal}{Journal of Physics D: Applied Physics}}
  \textbf{\bibinfo{volume}{40}}, \bibinfo{pages}{2666--2670}
  (\bibinfo{year}{2007}).

\bibitem{Thevenaz2008}
\bibinfo{author}{Th{\'{e}}venaz, L.}
\newblock \bibinfo{title}{{Slow and fast light in optical fibres}}.
\newblock \emph{\bibinfo{journal}{Nature Photonics}}
  \textbf{\bibinfo{volume}{2}}, \bibinfo{pages}{474--481}
  (\bibinfo{year}{2008}).

\bibitem{Okawachi2005}
\bibinfo{author}{Okawachi, Y.} \emph{et~al.}
\newblock \bibinfo{title}{{Tunable All-Optical Delays via Brillouin Slow Light
  in an Optical Fiber}}.
\newblock \emph{\bibinfo{journal}{Physical Review Letters}}
  \textbf{\bibinfo{volume}{94}}, \bibinfo{pages}{153902}
  (\bibinfo{year}{2005}).

\bibitem{Gersen2005}
\bibinfo{author}{Gersen, H.} \emph{et~al.}
\newblock \bibinfo{title}{{Real-space observation of ultraslow light in
  photonic crystal waveguides}}.
\newblock \emph{\bibinfo{journal}{Physical Review Letters}}
  \textbf{\bibinfo{volume}{94}}, \bibinfo{pages}{3--6} (\bibinfo{year}{2005}).

\bibitem{Khurgin2006}
\bibinfo{author}{Khurgin, J.~B.}
\newblock \bibinfo{title}{{Performance Limits on Delay Lines Based on Optical
  Amplifiers}}.
\newblock \emph{\bibinfo{journal}{Optics Letters}}
  \textbf{\bibinfo{volume}{31}}, \bibinfo{pages}{948--950}
  (\bibinfo{year}{2006}).

\bibitem{Khurgin2005}
\bibinfo{author}{Khurgin, J.~B.}
\newblock \bibinfo{title}{{Optical buffers based on slow light in
  electromagnetically induced transparent media and coupled resonator
  structures: comparative analysis}}.
\newblock \emph{\bibinfo{journal}{Journal of the Optical Society of America B}}
  \textbf{\bibinfo{volume}{22}}, \bibinfo{pages}{1062} (\bibinfo{year}{2005}).

\bibitem{Fan2015}
\bibinfo{author}{Fan, L.}, \bibinfo{author}{Fong, K.~Y.},
  \bibinfo{author}{Poot, M.} \& \bibinfo{author}{Tang, H.~X.}
\newblock \bibinfo{title}{{Cascaded optical transparency in multimode-cavity
  optomechanical systems.}}
\newblock \emph{\bibinfo{journal}{Nature Communications}}
  \textbf{\bibinfo{volume}{6}}, \bibinfo{pages}{5850} (\bibinfo{year}{2015}).

\bibitem{Miller2010a}
\bibinfo{author}{Miller, D.}
\newblock \bibinfo{title}{{Optical interconnects to electronic chips.}}
\newblock \emph{\bibinfo{journal}{Applied Optics}}
  \textbf{\bibinfo{volume}{49}}, \bibinfo{pages}{70} (\bibinfo{year}{2010}).

\bibitem{Alduino2007}
\bibinfo{author}{Alduino, A.} \& \bibinfo{author}{Paniccia, M.}
\newblock \bibinfo{title}{{Interconnects: Wiring electronics with light}}.
\newblock \emph{\bibinfo{journal}{Nature Photonics}}
  \textbf{\bibinfo{volume}{1}}, \bibinfo{pages}{153--155}
  (\bibinfo{year}{2007}).

\bibitem{Lee2008b}
\bibinfo{author}{Lee, B.~G.} \emph{et~al.}
\newblock \bibinfo{title}{{Ultrahigh-Bandwidth Silicon Photonic Nanowire
  Waveguides for On-Chip Networks}}.
\newblock \emph{\bibinfo{journal}{IEEE Photonics Technology Letters}}
  \textbf{\bibinfo{volume}{20}}, \bibinfo{pages}{398--400}
  (\bibinfo{year}{2008}).

\bibitem{Miller2009}
\bibinfo{author}{Miller, D. A.~B.}
\newblock \bibinfo{title}{{Device Requirement for Optical Interconnects to
  Silicon Chips}}.
\newblock \emph{\bibinfo{journal}{Proc. of IEEE Special Issue on Silicon
  Photonics}} \textbf{\bibinfo{volume}{97}}, \bibinfo{pages}{1166--1185}
  (\bibinfo{year}{2009}).

\bibitem{Madden2007}
\bibinfo{author}{Madden, S.~J.} \emph{et~al.}
\newblock \bibinfo{title}{{Long, low loss etched As(2)S(3) chalcogenide
  waveguides for all-optical signal regeneration.}}
\newblock \emph{\bibinfo{journal}{Optics Express}}
  \textbf{\bibinfo{volume}{15}}, \bibinfo{pages}{14414--14421}
  (\bibinfo{year}{2007}).

\end{thebibliography}

\begin{thebibliography}{1}
\expandafter\ifx\csname url\endcsname\relax
  \def\url#1{\texttt{#1}}\fi
\expandafter\ifx\csname urlprefix\endcsname\relax\def\urlprefix{URL }\fi
\providecommand{\bibinfo}[2]{#2}
\providecommand{\eprint}[2][]{\url{#2}}

\bibitem{Winful2013a}
\bibinfo{author}{Winful, H.}
\newblock \bibinfo{title}{{Chirped Brillouin dynamic gratings for storing and
  compressing light}}.
\newblock \emph{\bibinfo{journal}{Optics express}}
  \textbf{\bibinfo{volume}{21}}, \bibinfo{pages}{10039--10047}
  (\bibinfo{year}{2013}).

\bibitem{Boyd2003}
\bibinfo{author}{Boyd, R.~W.}
\newblock \emph{\bibinfo{title}{{Nonlinear Optics}}} (\bibinfo{publisher}{Acad.
  Press}, \bibinfo{year}{2003}).

\bibitem{Winful2015}
\bibinfo{author}{Dong, M.} \& \bibinfo{author}{Winful, H.~G.}
\newblock \bibinfo{title}{{Area dependence of chirped-pulse stimulated
  Brillouin scattering: implications for stored light and dynamic gratings}}.
\newblock \emph{\bibinfo{journal}{Journal of the Optical Society of America B}}
  \textbf{\bibinfo{volume}{32}}, \bibinfo{pages}{2514} (\bibinfo{year}{2015}).

\bibitem{McCall1969}
\bibinfo{author}{McCall, S.~L.} \& \bibinfo{author}{Hahn, E.~L.}
\newblock \bibinfo{title}{{Self-Induced Transparency}}.
\newblock \emph{\bibinfo{journal}{Physical Review}}
  \textbf{\bibinfo{volume}{183}}, \bibinfo{pages}{457--485}
  (\bibinfo{year}{1969}).

\bibitem{Sterke1991}
\bibinfo{author}{Sterke, C. M.~D.}, \bibinfo{author}{Jackson, K.~R.} \&
  \bibinfo{author}{Robert, B.~D.}
\newblock \bibinfo{title}{{Nonlinear coupled-mode equations on a finite
  interval: a numerical procedure}}.
\newblock \emph{\bibinfo{journal}{Journal of the Optical Society of America B}}
  \textbf{\bibinfo{volume}{8}}, \bibinfo{pages}{403--412}
  (\bibinfo{year}{1991}).

\end{thebibliography}
\end{document}